\documentstyle[11pt]{article}
\newcommand{\f}{\begin{equation}}
\newcommand{\ff}{\end{equation}}
\newcommand{\GG}{\alpha}
\newcommand{\blankline}{\vskip .3cm}

\begin{document}
\vfill
\centerline{\LARGE The physical hamiltonian}
\centerline{\LARGE in}
\centerline{\LARGE non-perturbative quantum gravity}
\rm
\vskip1cm
\centerline{Carlo Rovelli ${}^*$}
\blankline
\centerline{\it Department of Physics, University of
        Pittsburgh, Pittsburgh, Pa  15260; }
\centerline{\it Dipartimento di Fisica, Universit\'a di Trento;
INFN, Sez di Padova, Italia}
\blankline\blankline
 \centerline{Lee Smolin${}^\dagger$}
\blankline
 \centerline{\it   Department of Physics, Syracuse University,
 Syracuse, NY 13244; }
\centerline{\it Center for Gravitational Physics and Geometry,
Pennsylvania State University,}
\centerline{\it University Park, PA  16802-6360, USA}
 \vfill
\centerline{\today}
\vfill

\centerline{\bf Abstract}

A quantum Hamiltonian which evolves the gravitational field
according to time as measured by constant surfaces of a scalar
field is defined through
a regularization  procedure based  on the loop representation, and
is
shown to be finite and diffeomorphism invariant.  The problem of
constructing this Hamiltonian is reduced to a combinatorial and
algebraic problem  which  involves  the  rearrangements  of  lines
through
the vertices of arbitrary graphs.  This procedure also provides a
construction of the Hamiltonian constraint as a finite
operator on
the space of diffeomorphism invariant states as well as a
construction of the
operator corresponding to the spatial volume of the universe.

\blankline
\noindent
 \vfill
${}^*\ $ rovelli@pittvms.bitnet,
$\ \ {}^\dagger$  smolin@suhep.phy.syr.edu
\eject

One of the main problems of nonperturbative quantum gravity has
been how to realize physical time evolution in the absence of a
fixed background spacetime geometry \cite{time}.
One solution to this problem,
which has been often discussed, is to use a matter degree of
freedom to provide a physical
clock \cite{clock}, and represent evolution as
change with  respect it.   In  this letter  we  show  that  it  is
possible to
explicitly implement this proposal in the full theory of
quantum general relativity, using the nonperturbative
approach \cite{review,review-ls,abhay-leshouches}
based on the Ashtekar variables \cite{abhay}
and the loop representation \cite{carlolee,gambini}.
In particular, we use a scalar field as a clock as suggested
in several recent papers \cite{karel-sclock,ls-sclock},
and we show that it is possible to construct the hamiltonian
operator $\hat H$ that gives the evolution in this clock time.

We construct the hamiltonian operator $\hat H$ by using
regularization  techniques   recently  introduced   for  generally
covariant
theories \cite{review-ls,weaves,carloobserves,antisymm}.
The main technical result that we obtain is that the
operator $\hat H$, although constructed through a regularization
procedure that breaks diffeomorphism invariance, is nevertheless
diffeomorphism invariant, background independent and (as we have
argued elsewhere \cite{review,review-ls,carloobserves,antisymm}
is implied by these conditions) finite.   It follows that
$\hat H$ is well defined on the space $\cal V$ of the
spatially diffeomorphism invariant states of the
gravitational field.   Since this space $\cal V$, which we call the
knot space,
is spanned by the basis given by the generalized
knot classes\cite{carlolee}
(diffeomorphism equivalence classes
of finite sets of loops in $\Sigma$, the three-dimensional space
manifold), $\hat H$ is represented by an infinite
dimensional matrix in knot space.  We present here a procedure for
computing all the matrix elements of the hamiltonian
$\hat H$ in knot space.
This procedure is purely combinatorial and algebraic.
Thus, our main result is the reduction of the problem
of computing the physical evolution of the quantum gravitational
field with respect to a clock
to a problem in graph theory and combinatorics .

\blankline

We begin by introducing the scalar field $T(x)$, whose
three-surfaces of constant values may be taken, under appropriate
circumstances, to represent time \cite{karel-sclock,ls-sclock}.
If we denote its conjugate momentum by $\pi(x)$, the Hamiltonian
constraint has, classically, the form,
\f
{\cal C} (x) = {\scriptstyle{1 \over 2 \mu}}\ \pi^2 +
{\scriptstyle{\mu \over 2}} \ \tilde{\tilde{q}}^{ab}\
\partial_a T\ \partial_b T  + {\cal C}_{grav}
\ff
where $\mu$ is a constant with dimensions of energy density,
necessary so that $T$ have dimensions of time.
The gravitational contribution has the standard form \cite{abhay}
\f
{\cal     C}_{grav}      =   \epsilon_{ijk} \tilde{E}^{ai}
\tilde{E}^{bj}F_{ab}^k
+ \Lambda\  q = {\cal C}_{Einst} + \Lambda\ q
\ff
where $q=\det(q_{ab})$,  $\Lambda$ is  the cosmological  constant,
and
all other symbols have the usual meaning in the
Ashtekar formalism \cite{abhay}.    We then
restrict the freedom of choosing the time coordinate
by fixing the gauge $\partial_a T=0$.  It is not difficult to
show \cite{ls-sclock} that this implies that the lapse is
$N(x)=a /\pi(x)$ for some constant $a$ and that
all of  the infinite  number  of  hamiltonian  constraints  ${\cal
C}(x)$
turn out to be gauge fixed, except one, which is an integral over
the three manifold $\Sigma$:
\f
\int_\Sigma {{\cal C}\over\pi} = (2 \mu )^{-{1\over 2}}
\int_\Sigma \pi + \int_\Sigma
\sqrt{-{\cal C}_{grav}  } =0.
\ff
This gauge fixing commutes with the generators of spatial
diffeomorphisms, which may be imposed exactly
on the $T= constant$ surfaces.  If we go over to the quantum
theory the diffeomorphism invariant states are
then of  the form  $\Psi [  \{ \GG  \} ,  T ]$,  where $\{ \GG \}$
indicates
a generalized knot class, namely a diffeomorphism equivalence
class of (multiple) loops $\GG$, and the real number
$T$ is the constant value of the time.  These states then
satisfy a Schroedinger-like equation
\f
\imath \hbar\ {d\Psi \over dT } =\sqrt{2 \mu } \ \  \hat H \Psi
\ff
where $\hat H$ is the quantum operator corresponding to the
observable
\f
H = \int_\Sigma \sqrt{-{\cal C}_{grav}  }.
\label{hamiltonian}
\ff

We now proceed to construct the quantum operator $\hat H$.
We regularize
the integral  by writing it as a limit of a sum, and, in addition,
we
regularize each operator product according to the techniques
developed in refs.\cite{review-ls,weaves,carloobserves,antisymm}.
We write
\f
\hat{H}= \lim_{L  \rightarrow 0,A \rightarrow 0,\delta \rightarrow
0}\
\sum_I\ L^3\ \sqrt{ -\hat{\cal C}_{Einst \ I}^{L,\delta ,A}-
\Lambda \ \hat{q}_I^L }
\ff
where we have divided the spatial manifold $\Sigma$ into cubes
of size $L$ according to an arbitrary set of fixed euclidean
coordinates, and the sum is over these cubes, labeled $I$.  The
quantities  $\delta$ and $A$ are parameters involved in
the regularization  of the  Einstein term.  The order in which the
limits
have to  be taken  is a  crucial part  of the  definition  of  the
quantum
operator: we  specify this  order below.   The  Einstein  term  in
detail is
\begin{eqnarray}
\hat{\cal C}^{L,\delta,A}_{Einst \  I } = && {1 \over {2 L^3 A}}
\int_{cube\ I} d^3x \int d^3y \int d^3z\
f^\delta (x,y) f^\delta (x,z)
\sum_{\hat{a} < \hat{b}}   \nonumber \\  & &
\left [
\hat{T}^{\hat{a} \hat{b} }[ \gamma_{xyz} \circ
\gamma_{x\hat{a}\hat{b}}^{A} ](y,z) +
\hat{T}^{\hat{a} \hat{b} }[ \gamma_{xyz} \circ
\gamma_{x\hat{a}\hat{b}}^{A \ -1 } ](y,z)
\right ]
\label{definition}
\end{eqnarray}
where $ f^\delta (x,y) = {3 \over {4 \pi \delta^3 }}
\Theta ( \delta - | x-y| ) $ regulates the distributional
products ($\Theta$ is the step function).
$\hat T^{ab}$ is defined in \cite{review-ls,carlolee,weaves},
$\gamma_{xyz}$ is a zero area curve
running from $x$ to $y$ to $z$ and back and
$\gamma_{x \hat{a} \hat{b}}^{A} $ is a circle based at $x$
in the $\hat{a}\hat{b}$ plane with area $A$, all defined with
respect to the arbitrary euclidean coordinates.
This operator depends on three regularization scales, $\delta, A $
and $L$.
It is straightforward to check that the classical expression
corresponding to (6,7) reproduces (\ref{hamiltonian})
when the limits are taken.

Let us now study the action of (\ref{definition})
on a loop state $\Psi [\GG]$.
First, we notice that for fixed $L$ and $A$ the limit
$\delta \rightarrow 0$ is zero
unless there  is an  intersection or  kink in the $I$'th cube.  In
these
cases, let us label the $n$ lines going into the intersection
point, $p$, as $\alpha_i$, $i=1,...,n$.   We then have,
using the explicit forms of the operator $\hat T^{ab}$
\cite{review,review-ls,carlolee},
\begin{eqnarray}
\hat{\cal C}_{Einst \ I}^{L,\delta,A} \Psi [\GG ] =&&
{l_{Pl}^4 \over A L^3 } \sum_{i < j \leq n}
X(I,i,j,\delta )\  \sin (\theta_{ij} ) \sum_r
{\scriptstyle(-1)^r}
 \nonumber \\ &&
\left [
\Psi [ (\GG *_i\!*_j \
\gamma_{p\ \dot{\alpha}_i(p) \dot{\alpha}_j (p)}^{A})^r  ] +
\Psi [ ( \GG *_i\!*_j  \
\gamma_{p\ \dot{\alpha}_i(p) \dot{\alpha}_j (p)}^{A \ -1 })^r  ]
\right ]
\label{action}
\end{eqnarray}
where
\f
X(I,i,j,\delta )=\int_{cube\ I} d^3x \int ds \int dt\
f^\delta (x, \alpha_i (s))f^\delta (x, \alpha_j (t)),
\ff
$l_{Pl}$ is the Planck length, for simplicity we have
chosen a  parametrization such  that $|\dot{\alpha}^a_i  (s)|=1$,\
and
$*_i$ indicates the action of a "grasp" \cite{carlolee}
on the line $\alpha_i$. The angle
$\theta_{ij}$ is the angle between the $i$'th and $j$'th tangent
vectors at the intersection point $p$ and
$\gamma_{p\ \dot{\alpha}_i(p) \dot{\alpha}_j (p)}^{A} $ is a
loop based at $p$ in the $\dot{\alpha}_i(p) \dot{\alpha}_j (p)$
plane, with  area $A$.   This  loop, and  the $\sin(\theta_{ij} )$
factor,
appear because we have used the identity
\begin{eqnarray}
&& \sum_{\hat{a} < \hat{b}}
v^{\hat{a}}w^{\hat{b}} \left [
\Psi [  \GG *\!* \
\gamma_{x\ \hat{a}\hat{b}}^{A}  ] +
\Psi [ \GG *\!*  \
\gamma_{x\hat{a}\hat{b}}^{A \ -1 } ]
\right ] =
\nonumber \\
&& = |\vec v|  |\vec w| \sin ( \theta_{\vec v\vec w} )
\left [
\Psi [ \GG *\!*
\gamma_{x\ \vec v\vec w}^{A} ] +
\Psi [ \GG *\!*\gamma_{x\ \vec v\vec w}^{A \ -1 } ]
\right ]
\end{eqnarray}
which is  true for  every two vectors $\vec v$ and $\vec w$, with
an
angle  $\theta_{\vec v\vec w}$ between them,
because its classical counter\-part expressed in terms
of traces  of holo\-no\-mies  is  true,  to  order  $A$,  for  all
connections,
and we require that all such identities be satisfied in the
loop representation \cite{abhay-leshouches}.

An explicit  calculation then  shows that  for $\delta  \ll L$, we
have
$X(I,i,j,\delta )=$ $ 4\pi^2/\sin(\theta_{ij})\delta$, so that the
angular dependence in (\ref{action}) cancels. This cancellation
is the first "fortunate accident"  that makes it possible
to define a diffeomorphism invariant operator.

Assuming that  $L$ has been taken small enough so that there is at
most
one intersection per cube, we then have
\f
\hat{\cal C}_{Einst \ I}^\delta \Psi [\GG ] =
{4 \pi^2 l_{Pl}^4 \over \delta A L^3 }\ \
{\cal O}^A_I \Psi [\GG ] + O(\delta^2 /A) + O (\delta /L)
\ff
where the operator ${\cal O}^A_I$ is zero unless there is an
intersection in the box $I$, in which case it is given by
\begin{eqnarray}
{\cal O}^A_I \Psi [\GG ]   = &&
 \sum_{i<j \leq n} \sum_r {\scriptstyle(-1)^r}
 \nonumber \\  &&
\left[
\Psi [ (\GG *_i\!*_j
\gamma_{p \dot{\alpha}_i(p) \dot{\alpha}_j (p)}^{A})^r  ] +
\Psi [ ( \GG *_i\!*_j
\gamma_{p \dot{\alpha}_i(p) \dot{\alpha}_j (p)}^{A \ -1 })^r  ]
\right ]
\end{eqnarray}
and is sensitive only to diffeomorphism invariant features of the
intersection: the rooting of the lines through it
and the linear dependences of the tangent
vectors at the intersection.  The effect
of the operatoris to add a loop of area $A$ (measured by the
fictitious background metric) based at the intersection point
in each plane made by each pair of tangent vectors at the
intersection, and then sum over rearrangements
of the rooting through the intersection.

Let  us   now  discuss   the  finiteness  and  the  diffeomorphism
invariance of
the operator. As far as finiteness is concerned, the problem is to
show
that the limits $\delta \rightarrow 0$, $A\rightarrow 0$ and
$L \rightarrow 0$ can be taken, respecting the conditions assumed
namely $\delta \ll L$ and $\delta^2 \ll A$, so that the resulting
operator in (6) is finite.  We can accomplish this task if we pose
$L = \kappa \delta$ and $A=\kappa^3 \delta^2$ and take first
the limit   $\delta \rightarrow 0$  at fixed $\kappa$, followed by
the limit
$\kappa \rightarrow  \infty$. This  limit  exists  and  is  finite
because
up to  terms of  order $\kappa^{-1}$  the powers  of $\delta$  and
$\kappa$
coming from  the operator  and from  the volume  $L^3$ outside the
square
root in the sum (6) cancel. This cancellation is the second
"fortunate accident" that makes the present construction possible.

As the dependence on the regularization parameters
cancels, we may go on to discuss the
diffeomorphism invariance of the operator.
Let us assume that $\Psi [\GG ] $ is a diffeomorphism
invariant state,  so that  $\Psi [\GG  ] = \Psi [\phi \circ \GG ]$
for
all $\phi \in Diff_0 (\Sigma)$.
To zeroth order in $\kappa^{-1}$ and $\delta$,
the action of ${\cal O}^A_I$
is diffeomorphism covariant, in the sense that
${\cal O}^A_I \Psi[\phi \circ \alpha] =
{\cal O}^A_{\phi^{-1} \circ  I} \Psi[\GG] $,
the reason being the following.
For fixed  $\GG$,  and for small enough $\delta$
the action of the operator becomes independent
of $\delta$, because the added loop doesn't link any
other component of $\GG$. More precisely,
for each given $\GG$ and $\kappa$,there is a
$\delta_0$ such that for all $\delta < \delta_0$ there
is a diffeomorphism $\phi_\delta$ such that
${\cal O}^{(\kappa^3\delta^2)}_I \Psi [\GG ] =
{\cal O}^{(\kappa^3 \delta_0^2)}_I \Psi [\phi_\delta \circ \GG ]
+O(\kappa^{-1}) =
{\cal O}^{(\kappa^3 \delta_0^2)}_I \Psi [\GG ]
+O(\kappa^{-1})$; the last equality following from the
diffeomorphism invariance of $\Psi$.
It follows that on the diffeomorphism invariant states
the limit  exists trivially  because close  to $\delta=0$  we have
that
${\cal O}^{(\kappa^3 \delta^2)}_I\Psi [\GG ]$
is constant in $\delta$  (provided that the intersection
remains in the box as the box is scaled down).
Moreover, the only effect on
${\cal O}^{(\kappa^3 \delta^2)}_I\Psi [\GG ] $ of a
diffeomorphism on $\GG$
is, (up to errors of order $\delta$ and $\kappa^{-1}$), to
possibly take the intersection outside the box. This is because
in the  limit the  action of  the operator  (adding a  small loop,
which
doesn't link anything, in the planes defined by the pairs of
tangent vectors, and rearranging the rootings at the intersection)
is
well defined on the diffeomorphism
equivalence classes of loops. Thus, the effect of a diffeomorphism
on $\GG$ can be simply compensated by moving the box
accordingly.  The result is that we have defined an operator
which is different from zero only if the box $I$ contains an
intersection, is finite, and transforms covariantly under
diffeomorphisms.  From this, we obtain below a genuinely
diffeomorphism invariant operator simply by summing over
all the boxes.

Next, the determinant of the three metric is regulated as
\f
\hat{q}^L_I= {1 \over 10 L^6}
\sum_{\hat{a}\leq\hat{b}\leq\hat{c}}
\int_{I\hat a}  d^2S_a(\sigma_1)
\int_{I\hat b}  d^2S_b(\sigma_2)
\int_{I\hat c}  d^2S_c(\sigma_3)  \
\hat{T}^{abc}(\sigma_1 , \sigma_2 ,
\sigma_3 )
\ff
where the integrals are over the faces of the cube, which we
labelled as $I\hat a$,  summing both
front and  back, and   $d^2 S_a = \epsilon_{abc}d^2 S^{bc}$ is the
area
element of the $\hat{a}$'th face of the $I$'th cube\footnote{This
definition is different from the one given in
\cite{review-ls}, which is not diffeomorphism invariant.}. From
the fact that as $L \rightarrow 0$ we have
$T^{abc}(\sigma_1 , \sigma_2 ,\sigma_3 ) = \epsilon^{abc} q$,
the correct classical limit is assured.  At the same time, this
sum leads (see \cite{weaves})
to the diffeomorphism covariant quantum action
\f
\hat{q}^L_I \Psi [\GG ] = {l_{Pl}^6 \over L^6}
\sum_{\hat{a}\leq\hat{b}\leq\hat{c}}
\sum_{i,j,k} I[{I \hat{a}}, \alpha_i ]
I[{I \hat{b} }, \alpha_j ]  I[{I \hat{c}}, \alpha_k ]\ \
{\cal W}\ \Psi [\GG ] + O(L)
\ff
where $ I[{I \hat{a}}, \alpha_i ]$ is the intersection number
between the $\hat{a}$'th face of the $I$'th cube and the
$i$'th line coming into the intersection and ${\cal W} $
is the (diffeomorphism covariant) linear operator that rearranges
the rooting through the intersection according to the
grasp defined by $\hat T^{abc}$, and is zero if there is
no intersection in the box.

Let us now put these results together. If we define the operator
${\cal M}_I$ by
\f
L^6 \left [
\hat{\cal   C}_{Einst   \   I}^{(\kappa   \delta),\delta,(\kappa^3
\delta^2)}
   +   \Lambda\ \hat{q}_I^L \right ] \Psi [\GG ] =
{\cal M}_I
\Psi [\GG ]
+O(\kappa^{-1})  +O(\delta ),
\ff
then  we   have  found  that  when  the  $I$'th  box  contains  an
intersection
\begin{eqnarray}
&& {\cal M}_I = \nonumber \\ &&
= \lim_{\delta \rightarrow 0}
\left[ 4 \pi^2 l_{Pl}^4 \ \ {\cal O}^{\kappa^3
\delta^2 }_I  + l_{Pl}^6 \Lambda
\sum_{\hat{a}\leq\hat{b}\leq\hat{c}}
\sum_{i,j,k} {\scriptstyle I[{I \hat{a}}, \alpha_i ]
I[{I \hat{b}}, \alpha_j ]  I[{I \hat{c}}, \alpha_k ]}\
{\cal W}
\right]
\nonumber \\ &&
+ O(\kappa^{-1})
\label{sum}
\end{eqnarray}
To complete the definition of the operator $\hat{H}$ we have to
take the square root and then take the limits.    Since
the only non--vanishing terms in the sum in (6) come
when there is an intersection in the box,  the sum reduces
to a sum over the intersections of $\GG$. We label these
intersections with the index $i$.  This sum is now
genuinely diffeomorphism invariant.  For each term,
the square  root is  equal to the square root of ${\cal M}_i={\cal
M}_{I(i)}$,
where, for every $\delta$ and $\kappa$, $I(i)$ is the box in which
there is the intersection $i$, plus terms that vanish as $\delta
\rightarrow 0$ for all fixed $\kappa$.  The limit
$\delta \rightarrow 0$ may then be taken,
and, if $\Psi $ is a diffeomorphism invariant state, the result
is,  according   to  the   argument  given  above,  diffeomorphism
invariant
up to terms of order $\kappa^{-1}$.
The limit $\kappa \rightarrow \infty$ may then be
taken, and the result is the diffeomorphism invariant
operator
\f
\hat{H}\Psi[ \GG  ] = \sum_{i}\ \  [ {\cal M}_{i} ]^{1\over 2} \ \
\Psi [ \GG ],
\label{final}
\ff
where the  sum is  now over  all the  {\it intersections}  $i$  of
$\GG$.

The action indicated by (\ref{final}) is finite and
diffeomorphism invariant.  It remains
to describe the form of the operator ${\cal M}$ and the meaning of
its square root.  To do this it is convenient to choose a basis
for the diffeomorphism invariant bra states of the form
$<m_1, ....,  m_n; {\cal  K}_P; a_1,...,  a_n|$.  This refers to a graph
with $n$ intersections with $2 m_i, i=1,...,n$ lines entering each
one.  The discrete infinite dimensional index ${\cal K}_P$ labels
the knotting and linking of a nonintersecting link class
with $P$ ordered open ends (a tangle\cite{baez}),
which are joined to the intersections; here $P=\sum_i 2 m_i$.
Given a knot with $P$ open lines attached to $n$
intersection, we can still have diffeomorphism inequivalent knots,
by varying the rooting, and/or the linear dependences among the
tangent vectors, at each intersection.
These inequivalent knots span a subspace of the state space, which
we
denote as ${\cal S}[{\cal K}_P, m_1, ...., m_n]$. This subspace is
isomorphic to the tensor product of a linear space $V_i$
for every  intersection $i$.   $V_i$   is  spanned  by  the  basis
vectors labelled
by $a_i$. If there are four or less distinct tangent vectors
at the point of intersection $V_i$  is
finite dimensional, as the only diffeomorphism invariant
information is contained in the  rooting and
in the possible coincidences of three or more tangent vectors in
planes. However, if there are five or more distinct lines entering
the
intersection, the diffeomorphism equivalence classes form
a finite dimensional configuration space which in general is
not discrete\footnote{Contrary to what has
been implied by the authors previously.}, because in general
the  diffeomorphism   equivalence  classes   of  loops  {\it  with
intersections\ }
are also labelled by continuous parameters.

The operator ${\cal M}_i$ then has the form of an infinite
dimensional matrix
\begin{eqnarray}
&&
<m_1, ....,  m_n, {\cal  K}_P ,  a_1, ..., a_i, ..., a_n |\  {\cal
M}_i =
\nonumber \\ &&
=\ <m_1,  ...., \tilde  m_i, ...,  m_n, \tilde{\cal K}_{P}  , a_1,
...
\tilde  a_i,  ...,  a_n  |  \  \  [{\cal  M}_i  ]_{a_i  m_i  {\cal
K}_P}^{\tilde a_i
\tilde m_i \tilde{\cal K}_P}
\end{eqnarray}
The term  coming from  $\det(q)$ is  diagonal in the indices
$m_i$and
${\cal K}_P$, because it only
rearranges the lines at the intersection.
The Einstein term is non--vanishing only in the
first upper diagonal in these indices: more precisely, it has the
form    $\delta^{m_i+1}_{\tilde    m_i}    \    \    \delta^{{\cal
K}_{P+2}}_{\tilde
{\cal K}_P}$, times a matrix in the $a_i$ space.
This is because its action is to add a new loop,
not linking anything, for every pair of
the $2m_i$  tangent vectors,  the added  loop being  in the  plane
defined by the
two tangents, followed by the rearrangement
of the rooting through the intersection.
This structure gives ${\cal M}_i$ a relatively
simple "block diagonal and upper-diagonal" form
\f
 [{\cal M}_i ]_{a_i m_i {\cal K}_P}^{\tilde a_i
\tilde m_i \tilde{\cal K}_P}
=
\delta^{m_i}_{\tilde m_i} \ \ \delta^{{\cal K}_{P}}_{\tilde
{\cal K}_P}\ [{\cal M}_i^q]_{a_i}^{\tilde a_i} +
\delta^{m_i+1}_{\tilde m_i} \ \
\delta^{{\cal K}^{\cup}_{P+2}}_{\tilde
{\cal K}_P}
[{\cal M}_i^{Einst}]_{a_i}^{\tilde a_i}.
\ff
where ${\cal K}^{\cup}_{P+2}$ is the $P+2$ tangle that is gotten from
the $P$ tangle ${\cal K}$ by adding a simple loop which links nothing
else in ${\cal K}$.

The explicit computation of the blocks
$[{\cal M}_i^q]_{a_i}^{\tilde a_i} $ and
$[{\cal M}_i^{Einst}]_{a_i}^{\tilde a_i}$ is a tedious but
straightforward exercize in three dimensional geometry and
combinatorics, defined by the rearranging of rooting through
the intersection  produced by  the grasps  of loop operators $\hat
T^{ab}$ and
$\hat T^{abc}$. In the computation one must take
into account the standard spinor identities
of the  loop representation  \cite{carlolee}: if the definition is
consistent,
the operator should commute with these identities.  The remaining
problem is to compute the square root of a block diagonal
and upper-diagonal infinite dimensional matrix.  A technique
is under  development to  accomplish this  order by  order in  the
number
of lines coming into each intersection.

In summary, the main result reported in this work
is the discovery of a regularization procedure that
provides the definition of a finite and diffeomorphism
invariant physical-time-hamiltonian $\hat H$,
and reduces the problem of computing its matrix elements
to  an   algebraic  problem  in  three--dimensional  geometry  and
combinatorics.
If the construction developed here is consistent, this geometrical
action of the gravitational hamiltonian $\hat H$, which is just to
add loops
and rearrange  rootings at  the intersections of the knots states,
should
code the full content of the Einstein equations, in a
diffeomorphism invariant form\footnote{as the harmonic
oscillator equations of motion are coded in the number operator.}.

We close with some comments.  First,
the techniques described here also provide a finite expression for
the
(full) hamiltonian  constraint; in  fact, we  can define  a finite
operator that
corresponds to $H(f)=\int_\Sigma f \sqrt{{\cal C}}$ for any $f$.
This makes  it  possible  to  define  the  hamiltonian  constraint
directly
on the space of diffeomorphism invariant states.
Using this operator, one should recover
previous results \cite{carlolee,solutions} on the kernel of
the hamiltonian constraint.  Second, as the $A^{-1}$ in (7) is
cancelled against other factors, the limit taken in (6) does not
define an area derivative.  It would be of interest to investigate
the
space of loop functionals on which this limit is well defined;
unlike  the  space  of  area  differentiable  states,  this  space
includes
the diffeomorphism invariant states.  Third, we note
that with the Einstein term left out, our construction
provides a definition of the diffeomorphism invariant operator for
the volume of the universe.  Fourth, the Hamiltonian that we have
defined here  could be an important ingredient for determining the
reality
conditions, and hence the physical inner product \cite{review-ls}.
Finally, we  note that  the finiteness  of the  Hamiltonian is not
sufficient
for the evolution operator to be finite.
It is necessary to compute at least the second
order term in the
expansion in time
of the evolution operator,
to  see   whether  the   sums   over   virtual   states   converge
\cite{sidney}.

\blankline
\blankline

We thank Abhay Ashtekar, John Baez, Sidney Coleman, Louis Crane,
Karel Kuchar and Jost Zeegward for important comments. This work
has been supported by the NSF grants  PHY-90-12099, PHY-90-16733,
the NSF  United States-Italy  Cooperative Research  Grant  INT-88-
15209,
and by Syracuse University's research founds.
LS would like to thank Angela e Franco Rovelli, Paola Cesari
and Anna Pigozzo for hospitality during the course of this work.

\end{document}